\documentclass{aa61}  
\usepackage{graphicx}
\usepackage{subfigure}
\usepackage{float}
\usepackage{psfrag}
\usepackage{amsmath}
\usepackage{amssymb}
\usepackage{ulem}

\usepackage{txfonts}

\begin{document}

   \title{Non-parametric modeling of the intra-cluster gas using\\ 
           APEX-SZ bolometer imaging data}

   \subtitle{}

   \author{
     K. Basu\inst{1 \and2}\thanks{\email{kbasu@mpifr-bonn.mpg.de}} \and
     Y.-Y. Zhang\inst{2} \and 
     M. W. Sommer\inst{2 \and1} \and 
     A. N. Bender\inst{3} \and 
     F. Bertoldi\inst{2} \and
     M. Dobbs\inst{4} \and 
     H. Eckmiller\inst{2} \and
     N. W. Halverson\inst{3} \and 
     W. L. Holzapfel\inst{5} \and
     C. Horellou\inst{6} \and
     V. Jaritz\inst{2} \and
     D. Johansson\inst{6} \and
     B. Johnson\inst{5} \and
     J. Kennedy\inst{4} \and 
     R. Kneissl\inst{7} \and
     T. Lanting\inst{8} \and 
     A. T. Lee\inst{5 \and9} \and 
     J. Mehl\inst{10} \and 
     K. M. Menten\inst{1} \and 
     F. P. Navarrete\inst{1} \and
     F. Pacaud\inst{2} \and 
     C. L. Reichardt\inst{5} \and 
     T. H. Reiprich\inst{2} \and
     P. L. Richards\inst{5} \and
     D. Schwan\inst{5} \and
     B. Westbrook\inst{5}
   }

   \institute{
     Max Planck Institute for Radio Astronomy, 53121 Bonn, Germany
     \and
     Argelander Institute for Astronomy, Bonn University, 53121 Bonn, Germany
     \and
     Center for Astrophysics and Space Astronomy, University of Colorado,
     Boulder, CO, 80309, USA 
     \and
     Physics Department, McGill University, Montreal, Canada H2T 2Y8
     \and
     Department of Physics, University of California, Berkeley, CA, 94720, USA
     \and
     Onsala Space Observatory, Chalmers University of Technology, 43992
     Onsala, Sweden
     \and
     Joint ALMA Observatory, Las Condes, Santiago, Chile
     \and
     Schoold of Physics and Astronomy, Cardiff University, CF243YB, UK
     \and
     Lawrence Berkeley National Laboratory, Berkeley, CA, 94720, USA
     \and
     University of Chicago, 5640 South Ellis Avenue, Chicago, IL 60637, USA
   }

   \date{Received ............. /  
      Accepted .............}

  \abstract
   {}
   {We aim to demonstrate the usability of mm-wavelength imaging data obtained from the APEX-SZ bolometer array to derive the radial temperature profile of the hot intra-cluster gas out to radius $r_{500}$ and beyond. The goal is to study the physical properties of the intra-cluster gas by using a non-parametric de-projection method that is, aside from the assumption of spherical symmetry, free from modeling bias.}
   {We use publicly available X-ray spectroscopic-imaging data in the 0.7--2 keV energy band from the \textit{XMM-Newton} observatory and our Sunyaev-Zel'dovich Effect (SZE) imaging data from the APEX-SZ experiment at 150 GHz to de-project the density and temperature profiles for a well-studied relaxed cluster, Abell 2204. We derive the gas density, temperature and entropy profiles assuming spherical symmetry, and obtain the total mass profile under the assumption of hydrostatic equilibrium. For comparison with X-ray spectroscopic temperature models, a re-analysis of recent \textit{Chandra} observation is done with the latest calibration updates. We compare the results with that from an unrelaxed cluster, Abell 2163, to illustrate some differences between relaxed and merging systems.}
   {Using the non-parametric modeling, we demonstrate a decrease of gas temperature in the cluster outskirts, and also measure gas entropy profiles, both of which are done for the first time independently of X-ray spectroscopy using the SZE and X-ray imaging data. The gas entropy measurement in the central 100 kpc shows the usability of APEX-SZ data for inferring cluster dynamical states with this method. The contribution of the SZE systematic uncertainties in measuring $T_e$ at large radii is shown to be small compared to  \textit{XMM-Newton} and \textit{Chandra} systematic spectroscopic errors. The total mass profile obtained using the hydrostatic equilibrium assumption is in agreement with the published X-ray and weak lensing results; the upper limit on $M_{200}$ derived from the non-parametric method is consistent with the NFW model prediction from weak lensing analysis.}
   {}


   \titlerunning{Non-parametric ICM modeling from APEX-SZ imaging data}
   \authorrunning{Basu et al.}
   \maketitle
%

\section{Introduction}

Current cosmological models are built upon two complementary approaches of astronomical observation: the statistical study of the ensemble properties in a large sample of objects (i.e. from surveys) and the detailed analysis of the individual objects for gaining better understanding of the physical processes affecting those ensemble properties. This is particularly important in the study of galaxy clusters, where extraction of cosmological parameters from large survey samples (X-ray, optical, or in the radio/mm wavebands) relies critically on our understanding of different mass observables, which depends on the detailed physical processes affecting constituent gas and galaxies.

Accurately determining the thermodynamic state of the intra-cluster medium (ICM) out to a large radius is critical for understanding the link between cluster mass and observables. 
For over a decade, observations of the thermal Sunyaev-Zel'dovich Effect (tSZE, hereafter simply SZE; Sunyaev \& Zel'dovich 1970, Birkinshaw 1999) have been considered as a promising complement to X-ray observations for modeling the ICM in galaxy clusters, yet only recently has it been possible to make meaningful de-projections of gas temperature and density profiles using  SZE imaging data from multi-pixel bolometer arrays, in combination with X-ray data. The APEX-SZ experiment (Dobbs et al. 2006, Halverson et al. 2009) employs one of the first such powerful multi-pixel Transition-Edge Sensor (TES) bolometer cameras, and a joint analysis of the ICM properties using SZE and X-ray data has been presented by Nord et al. (2009, Hereafter NBP09) for the massive cluster Abell 2163.

In this paper we use the de-projection method used in NBP09 on the prototypical relaxed cluster Abell 2204. Our non-parametric analysis does not rely on any prior physical models in the construction of temperature and density profiles (apart from the assumption of spherical symmetry), hence the results are not based on parametric model fits. 
We measure the ICM entropy profile, as well as demonstrate the decrease of the ICM temperature in the cluster outskirts, first time from an SZE imaging data and independently from the X-ray spectroscopy. The derived ICM and cluster properties are compared with available X-ray and lensing results to highlight the level of accuracy of this independent method.

Joint SZE/X-ray de-projection analysis is expected to become a standard tool in the near future for understanding the ICM physical state, as large numbers of resolved SZE maps will be available from the new generation SZE experiments. Our  analysis assumes the gas to be in thermal equilibrium to model its physical properties, but presence of multi-phase ICM due to gas clumping will drive the electron temperature lower than the ion temperature in the electron-ion plasma (Evrard et al. 1996, Nagai et al. 2000). Recent hydro-simulations by Rudd \& Nagai (2009) have shown, with a limited sample of halo models, that this deviation is small (about 5\%) near $r_{200}$ for a relaxed cluster. Joint SZE/X-ray analysis using interferometric measurement of the SZE with OVRO/BIMA (Reese et al. 2002) has already shown that clumping effects are not large in the cluster interior (within $r_{500}$). Jia et al. (2008) have demonstrated the effect of the gas clumping on SZE and X-ray derived gas temperatures, and also found that these two quantities are in very good agreement within $r_{500}$ for the massive relaxed cluster RXC J2228.6$+$2036. But at large radii the gas should get clumpier, due to the onset of filamentary structures. One vital goal for sensitive imaging of the SZE signal using wide-field, multi-pixel bolometer cameras, and its combination with the X-ray and weak-lensing measurements, will be to provide an ultimate tool for measuring the gas clumping and thermodynamic state near the cluster virial radius, to give a dynamic view on the growth of clusters through accretion.

\subsection{Previous SZE/X-ray joint modeling}

Due to the unavailability of resolved SZE images most of previous SZE/X-ray joint analysis studies have been limited to analytical or numerically simulated cluster models with idealized noise properties. Zaroubi et al. (2001) considered a method for reconstructing the triaxial structure of clusters based on Fourier slice theorem and applied it to a set of cluster simulations. Lee \& Suto (2004) also considered de-projection method combining SZE and X-ray data and applied to analytical cluster models. Puchwein \& Bartelman (2006) have employed the Richardson-Lucy de-projection technique to reconstruct the ICM and probe the dynamical state of clusters from simulations, and Ameglio et al. (2007) used a joint SZE/X-ray likelihood function maximization using a Monte Carlo Markov Chain (MCMC) for a similar objective. 

Modeling ICM properties from real SZE observations has been limited mainly to isothermal $\beta$-models (Cavaliere \& Fusco-Femiano 1978).  Holzapfel et al. (1997), Hughes \& Birkinshaw (1998) used isothermal models to constrain the Hubble parameter from observations of the clusters Abell 2163 and CL 0016$+$16, respectively, and later  
Reese et al. (2002) extended this analysis to a sample of 18 clusters detected by OVRO/BIMA.  De Filippis et al. (2005) used published SZE decrement values and X-ray imaging data to constrain the triaxial structure of clusters using isothermal $\beta$-models. 
Zhang \& Wu (2000) similarly used the $\beta$-model to combine SZE and X-ray data to derive central gas temperature in clusters. A more detailed parametric modeling has been done by Mahdavi et al. (2007) for the cluster Abell 478, using simultaneous fits to the X-ray, lensing and SZE data assuming parametric models for dark matter, gas and stellar mass distribution, and hydrostatic equilibrium.

Yoshikawa \& Suto (1999) first used Abel's integral inversion technique, originally proposed by Silk \& White (1978), for a non-parametric reconstruction of radial density and temperature profiles using analytical and simulated cluster models. More recently Yuan et al. (2008) has extended this method for the most X-ray luminous cluster RXC J1347.5-1145 using published $\beta$-model fit values from SZE and X-ray measurements. 
Extrapolation of the density and temperature profiles to the cluster outskirts based on such parametric modeling can be problematic, in particular for clusters with a very peaked central emission such as RXC J1347.5-1145. Additionally, this cluster is considered to be a merging system (Cohen \& Kneib 2002) where the assumptions of spherical symmetry and hydrostatic equilibrium may not be valid. The nearest approach to non-parametric modeling was made by Kitayama et al. (2004) for the same cluster, RXC J1347.5-1145, using a beta-model density profile to fit the X-ray surface brightness and obtaining fitted temperature values separately in each radial bin from their SZE imaging data. The small extent of their SZE map (less than 2 arcmin) limited the temperature modeling again to the cluster core region. 

\subsection{Scope of the present work}

In this paper we apply the non-parametric ICM modeling based on Abel's integral inversion technique, as presented in NBP09, to the well studied and dynamically relaxed galaxy cluster Abell 2204  ($z=0.1523$, $L_{\mathrm{X}}=26.9\times 10^{44}$ $h_{50}^{-2}$ ~erg s$^{-1}$ in the $0.1-2.4$ keV band, $T_{\mathrm{X}} = 7.21 \pm 0.25$ keV; Reiprich \& B\"ohringer 2002). 
The only assumptions in this analysis are spherical symmetry for reconstructing temperature and density profiles, and hydrostatic equilibrium (HSE) for reconstructing the total mass profile. 
The primary aim is to confirm the validity of this method for modeling the ICM distribution and cluster mass -- and compare the results with those obtained from deep X-ray spectroscopic and weak lensing data -- in a cluster where the assumptions of spherical symmetry and HSE are generally accepted to be valid. 

We compute the \textit{Chandra} spectral temperature profile with the latest calibration updates and  compare it with the SZE-derived temperature profile. In contrast to the X-ray spectroscopic  measurements from \textit{Chandra}, the SZE-derived ICM temperature measurements near the cluster virial radius are constrained primarily by the statistical uncertainties in the SZE data. This fact demonstrates the potential for stacking the SZE signal of several relaxed clusters to put tighter constraints on the slope of the gas temperature profile in the cluster outskirts (Basu et al., in preparation). For a single cluster (Abell 2204), the depth in the APEX-SZ map allows us to model the temperature profile with meaningful errors up to $\sim80\%$ of the cluster virial radius (which we take to be $r_{200}$, the radius within which the mean total density is 200 times the critical density).

From density and temperature profiles we derive other physical properties like total gravitational mass, gas mass fraction and the gas entropy index. The total mass modeling provides a quantitative comparison with the published X-ray and lensing results. The modeling of the gas entropy profile from SZE/X-ray imaging data is a first, and we compare the central entropy values of two clusters with different morphologies, A2204 and A2163 (APEX-SZ analysis of the latter was presented in NBP09). This comparison shows how the gas entropy in the cluster core derived from SZE/X-ray joint modeling can be used to infer the dynamical state of clusters without the need for X-ray spectroscopy. A further comparison of the baryonic fraction of the ICM between A2204 and two other dynamically complex clusters detected by APEX-SZ (Bullet and A2163) illustrates a statistically significant difference of $f_{\mathrm{gas}}$ inside $r_{2500}$.

All the scientific results in this paper are computed from the radial profiles of two observables: the SZE temperature decrement at 150 GHz, and the X-ray surface brightness in the $0.7-2$ keV band of XMM-Newton. In \S2 and \S3, we describe the map making and radial profile extraction steps from the X-ray and SZE data, and discuss the different systematic uncertainties associated with each profile. \S4 describes Abel's integral inversion method and presents our primary results in the form of the radial density and temperature profiles. In \S5 we present the other derived quantities like gas entropy and the total cluster mass profiles, and list the conclusions in \S6.

We use the currently favored $\Lambda$CDM cosmology with the following parameters:  $\Omega_m=0.27$, $\Omega_{\Lambda}=1-\Omega_m=0.73$, and the Hubble constant $H_0=70$ km s$^{-1}$ Mpc$^{-1}$. At redshift of $z=0.1523$, the angular diameter distance of Abell 2204 is 541.6 Mpc. To put the radial profiles in perspective using the  characteristic cluster radii, we adopted the maximum likelihood NFW fit parameters from Corless et al. (2009), $M_{200} = 7.1\times 10^{14} M_{\odot}$ and $c=4.5$, which gives  $r_{200}=1.76$ Mpc ($11.2'$), $r_{500}=1.16$ Mpc ($7.3'$) and $r_{2500}=0.51$ Mpc ($3.2'$).

\section{Extraction of the X-ray surface brightness profile}

This section describes the basic data analysis steps for X-ray map making, and the method for extracting the radial profile. A brief description of the analysis method is provided below, refer to references for further details. 
We discuss the main source of the X-ray systematic error caused by particle background, that is incorporated in the analysis. 

\subsection{\textit{XMM-Newton} observation and data reduction}

\begin{figure}[t]
\centering
\includegraphics[width=6.5cm]{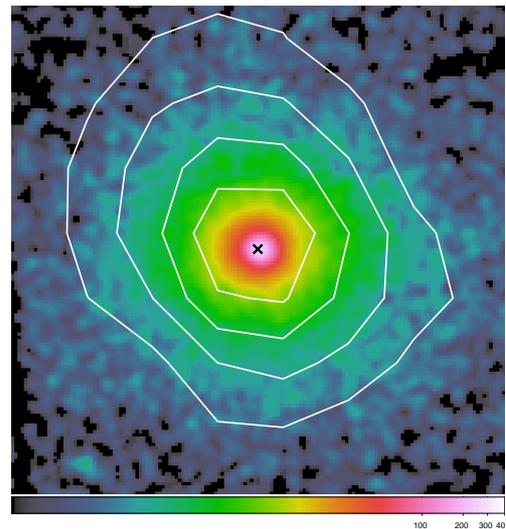}
\caption{$10\times 10$~arcmin \textit{XMM-Newton} MOS1 image of A2204, flat-fielded and smoothed with a $12^{\prime\prime}$ wide gaussian kernel. The overlaid contours (white) are from the APEX-SZ measurement;   
contour steps are -2, -4, -6 and -8 $\sigma$, and the SZ image resolution is 1 arcmin. 
The X-ray image is made in the 0.7-2~keV band, binned in $4^{\prime\prime}$ pixels. The color bar is in a logarithmic scale of $0.1-400$ counts per pixel. 
The black cross in the center denotes the flux weighted X-ray center used 
for the surface brightness profile extraction, and is within 4 arcsec of the SZE peak location obtained from spherical $\beta$-model fit.}
\label{fig:xmap}
\end{figure}

A2204 was observed by the \textit{XMM-Newton} EPIC camera with medium
filter in the full frame mode (ID: 0112230301). After carrying out the
screening procedure (e.g. Zhang et al. 2008, hereafter ZF08) to filter flares, we
obtained 17.5~ks, 18.5~ks and 14.3~ks clean exposure for the MOS1, MOS2
and pn instruments. For pn data, the fraction of the out-of-time (OOT)
events caused by read-out time delay is 6.30\%, and a simulated OOT event file is 
created to statistically correct for this. 
The SAS command ``edetect\_chain'' was
used to detect point-like sources, which were subtracted
before further data reduction. The vignetting correction to the 
effective area is accounted for by the weight column in the event
lists.  Geometric factors such as bad pixel and gap 
corrections are accounted for in the exposure maps. We choose the
\textit{XMM-Newton} blank sky accumulations in the Chandra Deep Field
South (CDFS) as background. The background observations were processed
in the same way as the cluster observations. The CDFS observations
used the thin filter, while the A2204 observations used 
the medium filter. The background of the A2204 observations is thus different from the CDFS using
the thin filter at energies below 0.7~keV. Therefore we performed all
the analysis at energies above 0.7~keV, in which the difference of the
background is negligible. The image of A2204 is shown in Fig.\ref{fig:xmap}.

\subsection{X-ray profile extraction}

\begin{figure}[t]
\centering
\includegraphics[width=8.5cm]{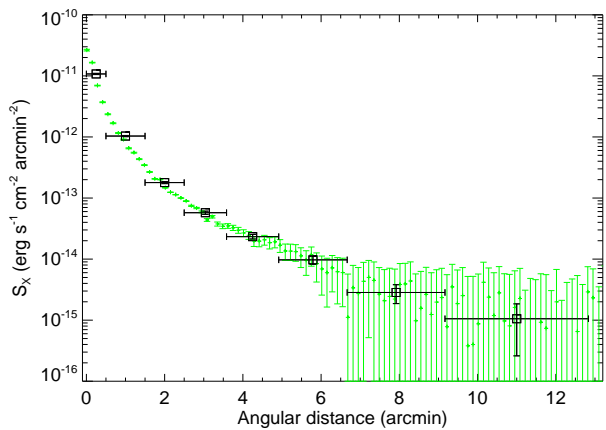}
\caption{X-ray surface brightness profile from the \textit{XMM-Newton} map in black squares, converted to physical units. The profile has been convolved and re-binned to match the APEX-SZ resolution. The errors include the systematic uncertainties due to background modeling and are incorporated in the de-projection analysis. Over-plotted in green are the surface brightness values prior to re-binning.}
\label{fig:xprofile}
\end{figure}

The 0.7--2 keV band is used to derive the surface brightness profiles.
This ensures an almost temperature-independent X-ray emission
coefficient over the expected temperature range. The width of the
radial bins is $2^{\prime \prime}$. An azimuthally averaged surface
brightness profile of the CDFS is derived in the same detector
coordinates as for the target.  
The count rate ratios of the target and
CDFS in the 10--12~keV band and 12--14~keV band for MOS and pn,
respectively, are used to scale the CDFS surface brightness.  
The residual background in each annulus of the surface brightness is the
count rate in the 0.7--2~keV band of the area scaled residual spectrum
obtained in the spectral analysis. Both the scaled CDFS surface
brightness profile and the residual background are subtracted from the
target surface brightness profile. The background subtracted and
vignetting corrected surface brightness profiles for three detectors
are added into a single profile, and re-binned to reach a significance
level of at least 3-$\sigma$ in each annulus out to $r\leq 9$ arcmin. 
The particle-induced background varies by less
than 10\% comparing the background observations.  Therefore the
dispersion of the re-normalization of the background observations is
typically 10\%. We take into account a 10\% uncertainty of the scaled
CDFS background and residual background. The resulting profile is shown in Fig.\ref{fig:xprofile}, and values are given in Table 1.

Note that raising the upper cut of the energy band does not provide
dramatic improvement in the signal-to-noise ratio in the surface
brightness profile in the cluster outskirts ($>3'$), where the gas
has a lower temperature and therefore does not contribute significant
X-ray photons at high energies. 

The X-ray surface brightness profile obtained from the above procedure
is convolved with a one-arcmin gaussian kernel to bring its
resolution to the same level as for the APEX-SZ raw image. This
smoothing raises the S/N ratio, particularly in the cluster outskirts. Additional re-binning is performed to conform the X-ray profile with the SZE data, since the latter is averaged in wide annular bins in the cluster outskirts to keep the statistical uncertainties under control. 
The widths of the radial bins are given in Table 1, where the central bin corresponds to the central 1 arcmin circle in each map. 
The resulting X-ray surface brightness profile after re-binning is shown
in Fig.\ref{fig:xprofile} (squares). The errors indicate 1~$\sigma$
uncertainty from the added poisson errors of the target and CDF
surface brightness profiles, \textit{plus} systematic uncertainties
due to the variation of the particle-induced background. Beyond
6~arcmin radius the systematic uncertainty starts to dominate, but we
still have a S/N of $\ge 3$ within 9~arcmin radius.  As we will
discuss in the next section, our results are currently dominated by the 
statistical uncertainties in the SZE imaging at this radius, therefore
we are not limited by X-ray systematics in the modeling of one single cluster.

\subsection{X-ray imaging vs. spectral spatial ranges}

Here we briefly highlight the advantage of the SZE/X-ray 
de-projection method to obtain the ICM temperature out to large radii, in
comparison with the X-ray spectral measurements (in particular from \textit{XMM-Newton} and \textit{Chandra}). To constrain the gas temperature to an uncertainty smaller than 10\%
from the X-ray spectra for such a hot cluster, one needs typically S/N
$>150$ in the 0.7--7.8~keV energy band (e.g. Zhang et al. 2009). Therefore such
temperature measurements are typically limited to the central regions
of clusters (up to a radius between $r_{2500}$ and $r_{500}$, less than half of the
cluster virial radius, see ZF08). 
Recently data from the \textit{Suzaku} satellite have been used 
to measure gas temperature beyond 
$r_{500}$ for a few clusters (Fujita et al. 
2008, Reiprich et al. 2009, George et al. 2009). 
However, these observations are expensive and limited to a few nearby ($z\lesssim 0.2$) 
clusters only. SZE/X-ray joint modeling can overcome this issue, by using 
X-ray surface brightness to provide primarily a constraint on 
the gas density, and then obtaining the temperature from SZE data. This 
easily allows for measuring the gas temperature at
the outer radii where the X-ray S/N is  low, e.g. 3--5. Thus ICM modeling up to the cluster
virial radius can be done, if the systematic uncertainties in both the X-ray and SZE imaging are controlled, and the SZE statistical uncertainties are brought down.

\section{Extraction of the SZE temperature decrement profile}

This section describes the basic reduction and map making steps for the
APEX-SZ data. The analysis is very similar to that of NBP09, which should be
consulted for further details. Here we emphasize the construction of a set
of SZE temperature decrement profiles, all consistent with our APEX-SZ
measurement, that we use to estimate the uncertainties in the de-projection analysis. 
A similar approach was also
used in NBP09, but the details of SZE profile construction and de-projection
procedure were not discussed.

\subsection{APEX-SZ observation and map making}

Abell 2204 was observed with the APEX-SZ camera in May 2008 and April 2009,
with roughly 80\% of the observing time spent in the 2008 run. The usable data
on the target amounts to approximately 10 hours, divided between scans of 20
minutes duration. The primary calibration source at 150 GHz was Mars, and
secondary calibrators were Neptune and RCW38. Details of the observing technique and data calibration
 for APEX-SZ are given in Halverson et al. (2009, hereafter HL09). We also refer to the Fig.1 of that paper for an illustration of the scanning pattern.

\begin{figure}[t]
\hspace*{-1.5cm}
\includegraphics[width=11cm]{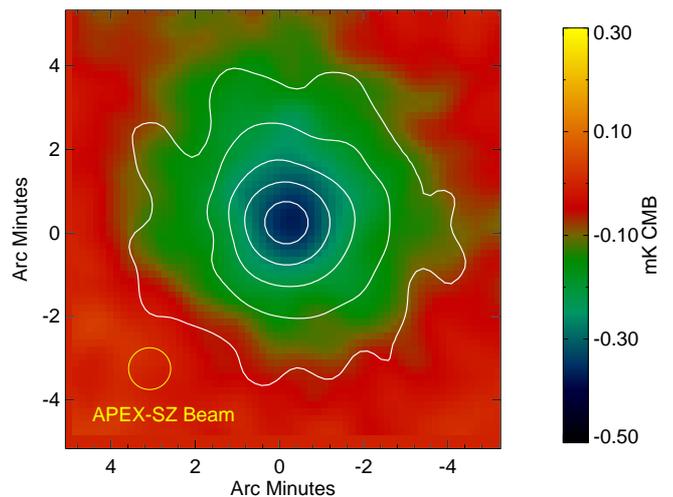}
\caption{$10\times 10$~arcmin APEX-SZ map of Abell 2204, overlaid with
  \textit{XMM-Newton} X-ray contours in steps of $3\sigma$ in logarithmic
  scale and smoothed to the APEX-SZ resolution. The SZE map has been
  deconvolved to the beam scale to reconstruct the full signal (see text). 
  The FWHM of the APEX-SZ beam is shown in the lower left. }
\label{fig:szmap}
\end{figure}

The reduction process is optimized for the circular drift scans employed for
the observation. After eliminating detectors with low optical response,
correlated atmospheric noise is removed by subtracting the
median signal across the good channels of the array at each time step after a temporary 
normalization step has been carried out. Additional reduction steps such as
despiking and de-glitching are used, but affect only a small amount of data.
Circular subscans are baselined, by defining the subscans consisting of 3 full
circles and then applying a fifth order polynomial, which corresponds to a low-pass
spatial filter (affecting spatial scales only marginally larger than those
filtered out by the circular scan pattern itself). For each scan, a map with
$15^{\prime\prime} \times 15^{\prime\prime}$ pixels is constructed, weighting the
data by the inverse rms at the position of each pixel in each scan. The result
we refer to as the ``raw map'', and the radial profile made from this map is
shown in Fig.\ref{fig:dcprof}.  In parallel, a bright point source
convolved with the instrument beam (obtained from fitting the Mars scans out
to a $4.5'$ radius) is processed by an identical pipeline to obtain the
transfer function (see HL09), which is used to perform the deconvolution.

The deconvolution of the map is performed iteratively in map space as
described by NBP09 and discussed in more detail by Nord (2009, PhD thesis).
The process essentially reconstructs the cluster signal as the sum of many
point sources as seen by the instrument beam. The final deconvolved map is
shown in Fig.\ref{fig:szmap}, overlaid with the X-ray surface brightness
contours. The noise on scales equal to the APEX-SZ beam is $44~\mu$K$_{\mathrm{CMB}}$ in the
central region of the map, corresponding to a peak signal-to-noise ratio of
$8.5$.

The outer contours of the APEX-SZ map with low
signal-to-noise ratio shows an elliptical shape. 
This is most likely the result of unfiltered noise
on scales of several arcminutes (but see Corless et al. 2009 for a discussion
on the triaxial dark matter halo in this cluster). We perform spherical and elliptical
isothermal $\beta-$model fits to the SZE map, which yield identical values for
the SZE emission center; (RA,Dec) $= (248.196, 5.577)$. These coordinates are
within 4 arcsec of the flux-weighted X-ray emission center, which is defined iteratively though a series of concentric circles in the X-ray map (see Zhang et al. 2010, \S 2.3). This 4 arcsec offset is comparable to
the pointing accuracy of the APEX telescope at 150 GHz. This provides additional
confirmation for the relaxed morphology of this cluster, to apply spherical de-projection using  a common SZE/X-ray center.

\subsection{SZE profile extraction and noise properties}

To estimate how uncertainties in the SZE map are propagated through our
analysis, we compute a set of (typically 100) deconvolved SZE profiles by
applying the above map making process on different noise realizations. The
resulting profiles are used to compute all the relevant cluster properties
(i.e.  profiles of gas density and temperature, and thereafter mass and
entropy profiles). 

\begin{figure}[t]
\centering
\includegraphics[width=8.5cm]{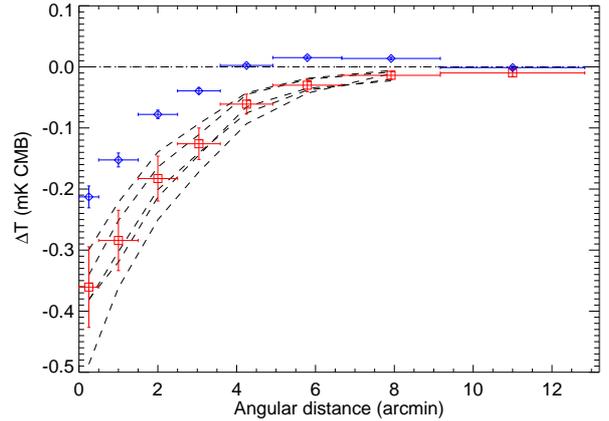}
\caption{Radial profile of the SZE temperature decrement at 150 GHz. The
  profile obtained from the raw reduction is shown by the blue diamonds, and
  the mean deconvolved profile by the red squares. The errors on the
  deconvolved profile represent the $1\sigma$ scatter in the set of profiles
  used in the de-projection analysis (see text). These errors are correlated
  especially within the narrow central bins. This reduces the significance of 
  the peak decrement as seen directly from the deconvolved map (Fig.\ref{fig:szmap}). 
  We show a random set of 5 deconvolved profiles (truncated at 8 arcmin) to
  illustrate the noise correlation.}
\label{fig:dcprof}
\end{figure}

We first obtain a set of ``jack-knived'' noise maps, by randomly selecting half
of the scan maps and inverting their signs, and then co-adding them with the
remaining scans (see HL09 for details of this technique applied to APEX-SZ
data). This removes all astrophysical signals but retains the noise structures
unfiltered by the pipeline. A random realization of the Cosmic Microwave Background (CMB) sky is
added to these noise maps to estimate the contamination of the SZE signal by
the CMB (\S3.3).  An azimuthally symmetric cluster map is made from the radial
profile of the raw map, and added to the
jack-knived noise maps. The thus simulated raw maps are deconvolved using the transfer function, and a
set of radial SZE profiles are obtained from the final maps. The scatter in this set of profiles constitutes the total statistical uncertainty in the SZE measurement, shown in Fig.\ref{fig:dcprof}.

The primary motivation for constructing a set of SZE radial profiles from jack-knived noise maps is to incorporate the effect of noise correlation occurring due to the presence of unfiltered noise structures in the map, which typically have scales much larger than the APEX-SZ beam. This reduces the significance of the
detection of the SZE signal, in particular if narrow binning is used. This is clearly seen from the errors on the final deconvolved radial profile in Fig.\ref{fig:dcprof}, which are about $50\%$ larger than the errors computed from the variance in each annular bin in the 
deconvolved cluster image (Fig.\ref{fig:szmap}). The total statistical uncertainties are then easily propagated through the de-projection analysis by computing the relevant physical quantities for each profile and measuring their scatter in each radial bin. 
This method also makes sure that the numerical errors coming from the de-projection method are not artificially enhanced  (see details in \S4.3).

Additionally, the choice of 1 arcmin binning in the central region of the SZE map leads to a correlation between the adjacent bins due to PSF smearing. The measured solid angle of the APEX-SZ beam is 1.5 arcmin$^2$, and $30\%$ of the beam power is in the near sidelobes outside the best fit gaussian beam with full-width at half maximum (FWHM) of $58^{\prime\prime}$ (Reichardt et al. 2009). Density and temperature de-projection based on a proper PSF deconvolution is not attempted in this paper; we simply note that the errors in the narrow central bins are possibly under-estimated by a small amount, with an overall downward bias in the measured gas temperature in these bins.  
The current choice of narrow binning inside $r_{2500}$ of the cluster is motivated by the aim of demonstrating
the compatibility of our temperature and mass profiles with published X-ray
results. It is also not desirable to smooth out the effect of the central cool
core of Abell 2204. 

The amount of correlation present between different radial bins is easily
computed by means of the correlation matrix. 
We compute the correlation matrix for the set of 100 deconvolved SZE profiles
and find that the 4 central bins are almost fully correlated ($\rho\gtrsim
0.8$), while the outermost bins have little correlation. For uniform binning
of the SZE profile (12 bins of 1 arcmin each) there is correlation between
adjacent bins all across the profile due to large scale noise structures, and in addition  
the bins near the center are correlated more strongly than the others 
due to PSF smearing. However, uniform binning is not used at large radii for extracting information out to the very low S/N regions of the SZE map, in a cluster whose detection significance is lower than those presented previously from APEX-SZ (HL09, NBP09). 
When fewer broad bins are used, the correlation becomes
negligible as can be expected ($\rho<0.1$ for 4 equal bins), but this is not used either as we are interested in the cluster cool core. 
This correlation pattern for any radial binning will propagate through all the other derived cluster quantities 
(like temperature, total mass and entropy bin values). Additionally, the noise will not go down as
expected when averaging several bins due to correlated errors, therefore we re-bin 
the original set of deconvolved maps for computing errors on averaged values.

\subsection{Sources of systematic errors}

The deconvolution method used in making the final SZE map can introduce
systematic bias in the final profiles. As in NBP09, a series of simulations is
performed by adding artificial cluster models ($\beta$-profiles) to jack-knived
noise maps and passing them through the reduction pipeline.  These are then
deconvolved using the transfer function, and the resulting profiles are
compared with the input $\beta$-models. The effect is a systematic lowering of
the cluster signal at large radii due to flux loss, by as much as $40\%$ at
$r_{200}$ (this number is true only if the real cluster profile follows an 
isothermal $\beta$-model). 
This error is considerably lower than the intrinsic
statistical uncertainties on the profile, which in case of A2204 is almost $100\%$
at $r_{200}$. Resulting systematic uncertainties on the ICM temperature profile are discussed in \S4.3, after describing the de-projection method.

For a large cluster like A2204 (virial radius $\sim 12$ arcmin), the
temperature anisotropies in the CMB are a major source of confusion. Following
the same prescription as in NBP09, we attempt to quantify this by making
multiple realizations of the CMB sky using the HEALpix software (Gorski et al.
2005), and adding these to the jack-knived noise maps before performing noise
simulations. The additional scatter in the resulting radial profiles is 14\%
at $r_{500}$ and roughly twice as large at $r_{200}$, again less than the
statistical errors in the APEX-SZ measurements. The systematic uncertainty on
the APEX-SZ measurement arising from calibration errors is of the order of 5\%
(HL09).

Other systematic errors in the SZ map can arise from unresolved point sources (radio or sub-millimeter galaxies) and galactic dust emission, which we have ignored. There is no indication of any point like sources in the 150 GHz SZE map. The NVSS radio catalog (Condon et al. 1998) lists a 70 mJy radio source at 1.4 GHz approximately $9^{\prime\prime}$ from the cluster X-ray center. After subtracting the best fit $\beta$-model from the raw map the rms noise at the map center is 2.2 mJy/beam, and no indication of a point source is seen in the residual raw map. The IR luminosity of the central brightest cluster galaxy (BCG) in Abell 2204 is reported by Quillen et al. (2008). The corresponding dust thermal emission at 150 GHz ($< 0.1$ mJy/beam) is much below the noise level at the map center, and the downward bias in the measured gas temperature at the cluster center can be ignored. Knudsen et al. (2008) found a bright sub-millimeter galaxy, SMM J163244.7$+$053452, in the field of A2204 at a distance of 39 arcsec N-W from the cluster center. Its $850 ~\mu$m flux density is $22.2\pm 4.9$ mJy, with estimated magnification of $\mu=3.4$. Assuming a spectral index $\alpha=3$, where $S_{\nu}\propto \nu^{\alpha}$, this source will produce a flux density of roughly $1.7$ mJy at 150 GHz, corresponding to a temperature increment of $34$ $\mu$K for the APEX-SZ beam. This is lower than the noise rms at the position of this galaxy in the map.

\section{De-projection of radial density and temperatures}

The three-dimensional (de-projected) density and temperature profiles are obtained directly using Abel's integral inversion method (as in NBP09), with the assumption of spherical symmetry. Although proposed nearly three decades ago for joint SZE/X-ray analysis (Silk \& White 1978), this method has remained largely unused. One possible reason for this limited application might be due to its numerical instability, as it involves computing derivatives at each point on the observed profiles. We have utilized the noise correlation in the real SZE data to partially overcome this problem, which makes Abel's inversion technique a particularly simple and intuitive method for de-projection. Unlike the standard ``onion-skin'' method of de-projection used in X-ray spectral analyses (Kriss et al. 1983), Abel's inversion is not dependent on the choice of the outermost bin. The strong anti-correlation in the de-projected temperature values between adjacent bins, a numerical artifact found in several geometrical de-projection techniques (see Ameglio et al. 2007), is also not significant.

\subsection{Method for de-projection}

For the de-projection analysis, the SZE temperature decrement can be written as the integral of the electron pressure along the line of sight as
\begin{equation}
\Delta T(R) = 2 A_{\mathrm{SZE}} 
\int_R^{\infty} g(x,T_e)\ n_e(r)\ 
T_e(r)\ \frac{r\mathrm{d}r}{\sqrt{r^2 - R^2}}
\label{eq:szrad}
\end{equation} 
where $A_{\mathrm{SZE}} = T_{\mathrm{CMB}}\ (k\sigma_{\mathrm{T}}/m_ec^2)$, $r$ is the physical radius
from the cluster center, $R=D_A \theta$ where $\theta$ is the projected
angular distance on the sky and $D_A$ is the angular diameter distance. $
T_e(r)$ and $n_e(r)$ are the electron gas
temperature and density radial profiles. $g(x,T_e)$ is the frequency dependence of the SZ signal, in which the gas temperature dependent relativistic correction terms have a small contribution ($\sim 5\%$ at 150 GHz for a 10 keV cluster). Therefore, we can neglect the radial temperature dependence in $g(x,T_e)$ and incorporate 
a fixed temperature value $g(x,T_e$=$8.26~$keV$)$ into the $A_{\mathrm{SZE}}$ factor, following the mean X-ray temperature from Arnaud et al. (2005). Note that we are ignoring any contribution from the kinematic Sunyaev-Zel'dovich effect (kSZE), as its contribution is likely to be much less than our imaging uncertainty.

In a similar way, the X-ray surface brightness profile can be
written as
\begin{equation}
S_{\mathrm{X}}(R) = \frac{2}{4\pi (1+z)^4}\ 
\int_R^{\infty} n_e^2(r)\ \Lambda_H(T_e(r))\  \frac{r\mathrm{d}r}{\sqrt{r^2 - R^2}}.
\label{eq:xrad}
\end{equation}
We compute the value of the X-ray emissivity function $\Lambda_H(T_e)$ in each radial bin with the MEKAL code in XSPEC (Mewe et al. 1982, Kaastra 1992), using models for metallicity and temperature radial profiles obtained from the spectral measurements of ZF08. The actual measured metallicity values within 0-3 arcmin radius changes from $0.5~Z_{\odot}$ to $0.3~Z_{\odot}$, corresponding change in $\Lambda_H(r)$ is 12\%. The weak temperature dependence of the soft X-ray emission in the $0.7-2$ keV energy band makes our results practically insensitive to any temperature model used in calculating $\Lambda_H(T_e)$, as noted in NBP09.

Using Abel's integral equation, equations (\ref{eq:szrad}) and (\ref{eq:xrad})
can be inverted to obtain joint radial density and temperature profiles (Yoshikawa \& Suto 1999) 
\begin{equation}
T_e(r)\ n_e(r) = \frac{1}{\pi A_{\mathrm{SZE}}}\ 
\int_{\infty}^r\ \dfrac{\mathrm{d}\Delta T(R)}{\mathrm{d}R} \ 
\dfrac{\mathrm{d}R}{\sqrt{R^2-r^2}};
\label{eq:szabel}
\end{equation} 
\begin{equation}
\Lambda_{H}(T_e(r))\ n^2_e(r)\  = 4(1+z)^4\ 
\int_{\infty}^r\ \dfrac{\mathrm{d}S_{\mathrm{X}}(R)}{\mathrm{d}R} \ 
\dfrac{\mathrm{d}R}{\sqrt{R^2-r^2}}.
\label{eq:xabel}
\end{equation} 
Equations (\ref{eq:szabel}) and (\ref{eq:xabel}) are integrated numerically by summing in radial bins from
$i_{\mathrm{min}}$ to $i_{\mathrm{max}}$, where $i_{\mathrm{max}}$ is the
index for the outermost bin, and $i_{\mathrm{min}}$ corresponds to $r/D_A$.

To show that our analysis results do not depend on any \textsl{a priori} knowledge of the radial temperature profile, we tried two alternative approaches for the computation of the emissivity function in addition to the MEKAL model. We used a mean value of the X-ray temperature in all bins to compute $\Lambda_H(T_e(r)=T_X)$, where $T_X=8.26$ keV (Arnaud et al. 2005). Alternatively, we used a weak power-law dependence of the emissivity function on the gas temperature, as $\Lambda_H(T_e(r)) \propto T_e(r)^{-1/6}$. The second assumption gives excellent approximation to the X-ray emissivity values from the MEKAL code if we assume the bulk of the cluster gas has a temperature in the range 2--14 keV. The de-projected ICM density and temperature profiles from all three methods, after combining with the SZE radial profiles, are essentially identical given the statistical error in each radial bin. This confirms the fact that the use of the soft-band X-ray data in our analysis is primarily providing the constraints on gas density profile, whereas the gas temperature constraints come from the SZE measurement.

\subsection{Radial profiles for gas density and temperature}

\begin{figure}[t]
\centering
\hspace*{2mm}
\includegraphics[width=8.5cm]{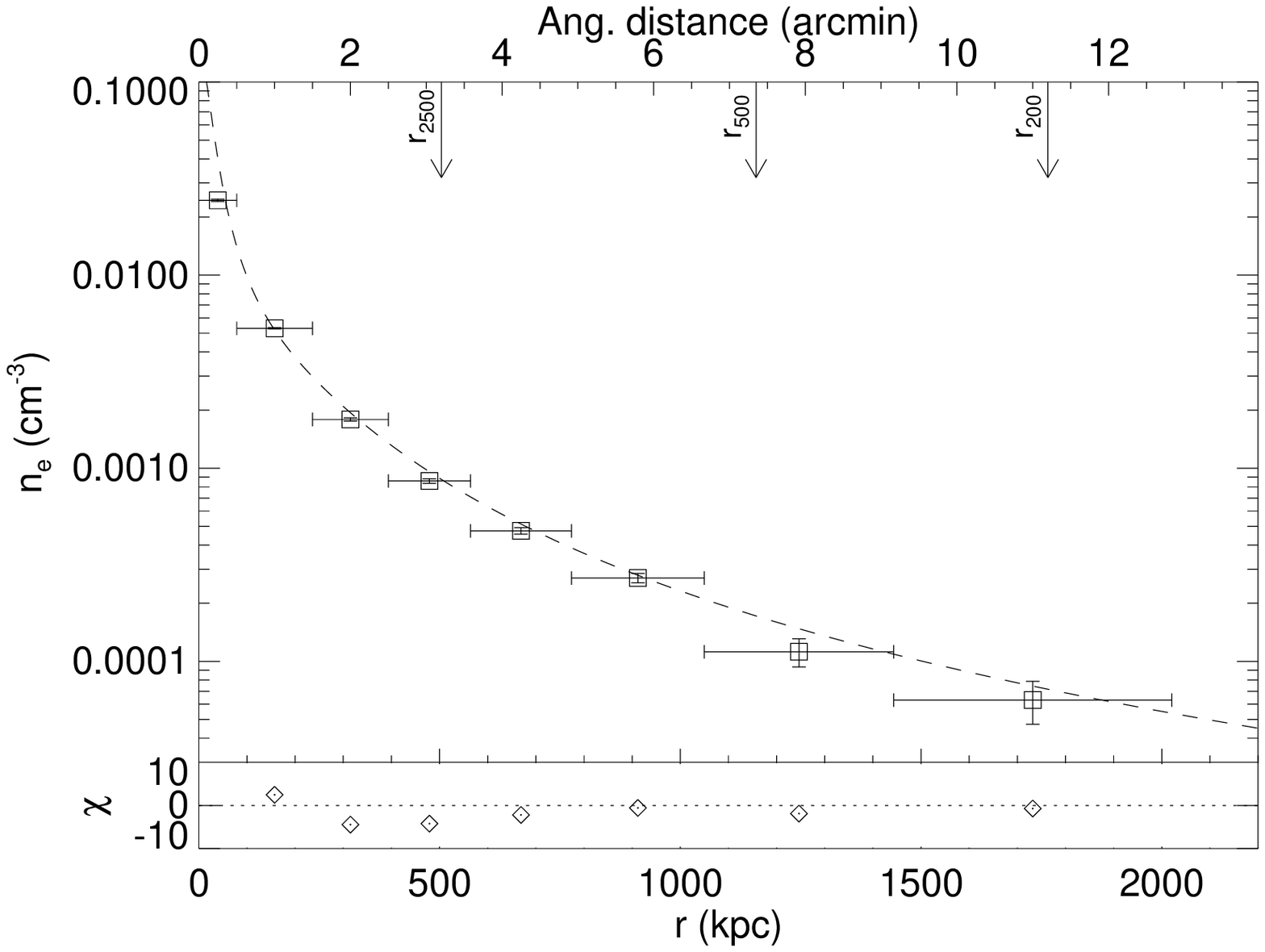}
\vspace*{2mm}

\includegraphics[width=8.5cm]{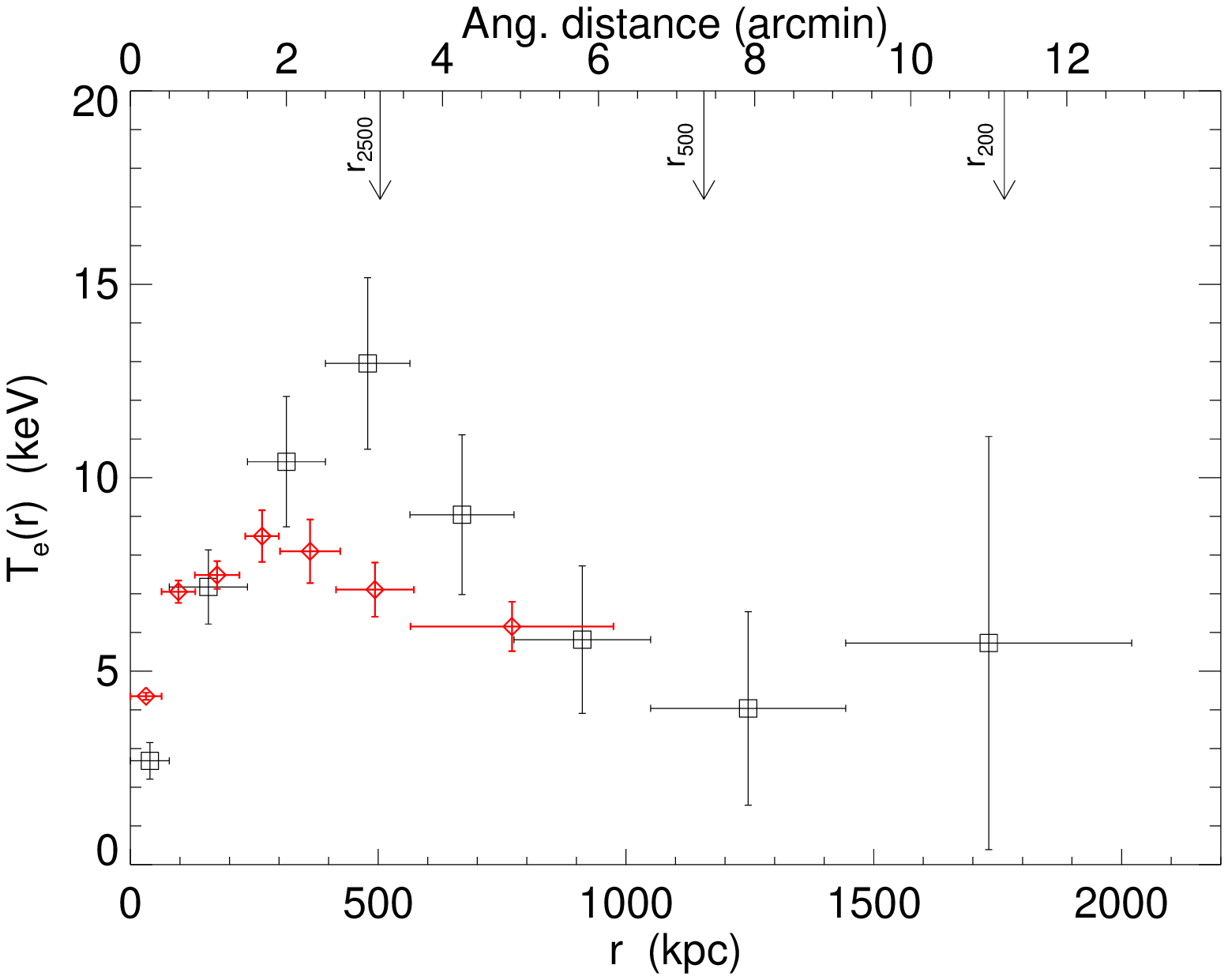}
\caption{\textit{Top panel:} The de-projected electron density with corresponding errors. The dashed line is the density profile from ZF08, obtained from fitting a double $\beta$-model to the X-ray surface brightness. The rms fractional differences from the $\beta$-model fit are shown below. 
\textit{Bottom panel:} The de-projected temperature values and their $1\sigma$ statistical errors from the SZE measurement. Over-plotted data points in red diamonds are from \textit{XMM-Newton} analysis by ZF08 (see \S4.4). The arrows in both plots mark the characteristic cluster radii, and the horizontal ``error bars'' the width of the bins.}
\label{fig:denstemp}
\end{figure}

The results for de-projection of density and temperature profiles for A2204 are shown in Fig.\ref{fig:denstemp}, and the corresponding values with their errors are  given in Table 1. 
Note that the uncertainties on the X-ray surface brightness profile due to the background modeling are included in the results, but the additional (small) systematic uncertainties from X-ray flux calibration are neglected, which likely produces an under-estimation of the errors on the density values in the inner bins. 
In the upper panel of Fig.\ref{fig:denstemp}, we overplot the density profile obtained by ZF08 by fitting a double $\beta$-model to the X-ray surface brightness. This density profile is \textit{XMM-Newton} PSF corrected, and the common $\beta$ slope parameter is obtained by fitting the outer component (see A.3 in ZF08). 
The rms fractional errors shown below Fig.\ref{fig:denstemp} are computed as $\chi = (\mathrm{bin~density}-\mathrm{model~density})/\mathrm{bin~error}$. Except for the inner arcminute where the X-ray brightness profile is extremely cuspy, the double beta model provides a good fit to our de-projected bin densities. This follows from the fact that in the 0.7--2 keV energy band the X-ray surface brightness is practically independent of the gas temperature. 
A similar argument had been used by Kitayama et al. (2004) while modeling the gas density profile with the X-ray derived $\beta$-model to obtain best fit radial temperature values in RXC J1347-1145. 

In the lower panel of Fig.\ref{fig:denstemp} the radial temperature profile is shown. There is a clear indication of the cluster cool core from APEX-SZ data; the temperature drops almost by a factor 3 from 500 kpc to 100 kpc radius. This is in contrast with the temperature profile for A2163 in NBP09, which could be fitted with a single isothermal profile at all radii within the $1\sigma$ uncertainties of the bin values. There is also a strong indication of a decreasing ICM temperature beyond its peak at $\sim500$ kpc. However, the temperature value at the last radial bin at $r_{200}$ is essentially an upper limit, there is no statistically significant SZE signal at this radius given the current noise level in the map.

\begin{figure}[t]
\centering
\includegraphics[width=8.5cm]{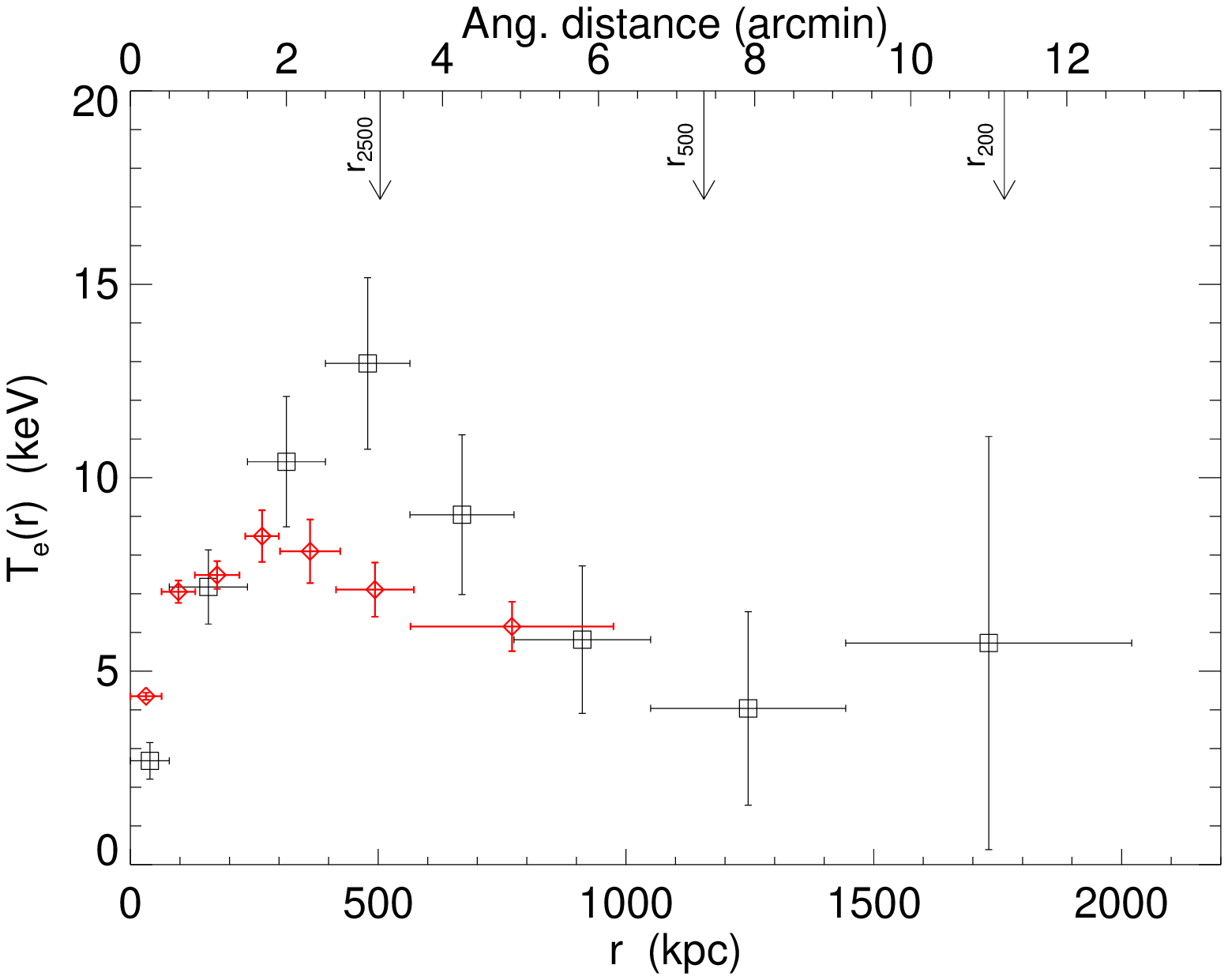}
\caption{The APEX-SZ measurement of the outer temperature profile in the cluster A2204, as compared to simulations and X-ray spectral measurements. The two data points represent the SZE-derived temperature values with $1\sigma$ errors; a decrease in gas temperature is supported at 98\% confidence level. The red dashed region represents a fit using UTP profiles from cluster simulations by Hallman et al. (2007), with $1\sigma$ errors.  The gray shaded area shows the average profile of cooling core clusters from ASCA (Markevitch et al. 1998), and the blue solid line is the \textit{Chandra} observation of cooling core systems (Vikhlinin et al. 2005).}
\label{fig:tempslope}
\end{figure}

The decreasing temperature profile in the cluster outskirts can be made clearer by re-binning the APEX-SZ data. As noted in \S3.2, the errors in the bin values are correlated and do not average down as expected in  random gaussian noise. Therefore, we re-bin the original set of deconvolved maps to compute the bin errors. The result is shown in Fig.\ref{fig:tempslope}, where we have divided the data in only two bins, excluding the central $3^{\prime}$ of the map. A decrease in gas temperature from its peak value is supported at $98\%$ confidence level ($2.3\sigma$). To put this temperature slope in perspective, we overplot in Fig.\ref{fig:tempslope} the results from recent X-ray observations and numerical simulations of clusters, scaled to the values for A2204. The solid line is the mean spectroscopic temperature profile in cooling core clusters (Vikhlinin et al. 2005), and the gray shaded region is the average profile of the cooling core clusters from ASCA with their $1\sigma$ dispersion (Markevitch et al. 1998). The SZE radial temperature is statistically consistent with both these measurements, although it appears to indicate a steeper slope. The SZE-derived temperature slope also appears steeper than the Universal Temperature Profile (UTP) fit from numerical simulations of relaxed cluster (Hallman et al. 2007), shown in the dashed line with the hatched region for the $1\sigma$ uncertainties in the UTP fit values. Again, SZE measurement from one cluster is not yet adequate to provide a quantitative comparison with the numerical simulations, but a stacking analysis of several relaxed clusters can be expected to yield a meaningful comparison by lowering the statistical noise.

As mentioned in the introduction, the two quantities $T_e$ and $T_{\mathrm{gas}}$ are used synonymously in this paper, where the latter is defined as $T_{\mathrm{gas}} = (n_e T_e + n_i T_i)/(n_e + n_i)$. They can differ if the post-shock equilibrium timescale between the electrons and ions is large, especially in the low density environment near $r_{200}$ (where $n_e \sim 10^{-4}-10^{-5} ~\mathrm{cm}^{-3}$). Recently, Rudd \& Nagai (2009) have provided quantitative estimate for this bias from cosmological hydrodynamic simulations of galaxy clusters, and found that for relaxed clusters (CL 104 in their simulations, with $T_{\mathrm{gas}}=5.4$ keV) $T_e$ can under-estimate the $T_{\mathrm{gas}}$ by about $5\%$ at $r_{200}$. Therefore this effect can be ignored for our current analysis of a single cluster. Their results most likely present the upper limit of this bias, since non-adiabatic heating due to shocks in the cluster outskirts is neglected in their models.

\begin{table*}
 \begin{minipage}[t]{\textwidth}
 \centering
 \caption{The input X-ray surface brightness and SZE temperature decrement values at each radial bin, and the different de-projected quantities derived from these two observables. The errors shown are the $1\sigma$ statistical uncertainties, including the systematic errors from the X-ray background modeling (the latter is included in the X-ray surface brightness errors). The first bin represents the central arcmin of the maps, corresponding to the APEX-SZ beam FWHM.}

 \begin{small}
 \begin{tabular}{cccccccc} \hline
 Bin & Radius & S$_{\mathrm{X}}$ & $\Delta T_{\mathrm{SZ}}$ & $n_e(r)$ & $T_e(r)$ & 
 Entropy\ \footnote[1]{entropy index, defined as $K=k_{\mathrm{B}}T_e n_e^{-2/3}$} & 
 $M_{\mathrm{tot}}(<r)$ \\
  & (arcmin) & (\ \footnote[2]{in units of $10^{-12}$ erg s$^{-1}$ cm$^{-2}$ arcmin$^{-1}$ }\ ) & 
 (mK) & ($10^{-3}$ cm$^{-3}$) & (keV) & (keV cm$^{2/3}$) 
 & ($10^{14}\ M_{\odot}$) \\ \hline\hline
  &  &  &  &  &  &  &  \\ 

1 & $0-0.5$ & $10.8\pm 0.2$ & $-0.36\pm 0.06$ & $24.4\pm 0.2$ & 
  $2.7\pm 0.5$ & $32\pm 6$ & $0.03\pm 0.01$ \\

2 & $0.5-1.5$ & $1.03\pm 0.02$ & $-0.28\pm 0.05$ & $5.30\pm 0.04$ & 
  $7.2\pm 0.9$ & $240\pm 30$ & $0.54\pm 0.04$ \\

3 & $1.5-2.5$ & $0.18\pm 0.007$ & $-0.18\pm 0.04$ & $1.79\pm 0.03$ &
  $10.4\pm 1.7$ & $710\pm 120$ & $1.72\pm 0.12$ \\

4 & $2.5-3.6$ & $0.06\pm 0.003$ & $-0.13\pm 0.03$ & $0.86\pm 0.02$ & 
  $12.9\pm 2.2$ & $1430\pm 270$ & $2.96\pm 0.41$ \\

5 & $3.6-4.9$ & $0.02\pm 0.002$ & $-0.06\pm 0.02$ & $0.47\pm 0.02$ &
  $9.0\pm 2.1$ & $1490\pm 370$ & $3.82\pm 0.83$ \\

6 & $4.9-6.7$ & $0.01\pm 0.001$ & $-0.03\pm 0.01$ & $0.27\pm 0.02$  &
  $5.8\pm 1.9$ & $1390\pm 510$ & $3.64\pm 0.61$ \\

7 & $6.7-9.2$ & $(2.8\pm 1.0)\times 10^{-3}$ & $-0.014\pm 0.007$ & $0.11\pm 0.02$ &
  $4.0\pm 2.5$ & $1700\pm 1200$ & $3.9\pm 1.3$ \\

8 & $9.2-12.8$ & $(1.0\pm 0.8)\times 10^{-3}$ & $-0.010\pm 0.006$ & $0.06\pm 0.02$ &
  $5.7\pm 5.3$ & $< 7500\ ^{\mathit{c}}$ & $< 13.1$\ \footnote[3]{upper limit at 68\% confidence level} \\

  &  &  &  &  &  &  &  \\
 \end{tabular}
 \end{small}

 \end{minipage}
\label{radtable}
\end{table*}

\subsection{Errors on the de-projected profiles}

As seen from Eqns.(\ref{eq:szabel}) and (\ref{eq:xabel}), the de-projection of density and temperature requires taking derivatives at each radial bin, which are the major source of introducing modeling errors onto the de-projected profiles. This fact may possibly have limited the application of Abel's inversion in the SZE simulations, using realistic mock observations with white noise. 
Although the high S/N imaging data from APEX-SZ with 1 arcmin resolution makes the application of Abel's inversion method feasible for the first time, propagating errors through a Monte-Carlo method will lead to a high and unphysical error level if the noise correlations between radial bins are ignored. 
As noted by Yoshikawa \& Suto (1999), pre-smoothing the data will reduce this error, but due to its model dependent nature we refrain from smoothing. It is also difficult to determine the degree of smoothing to be applied: a small smoothing kernel is insufficient to offset the numerical error (in particular for the narrow central bins), whereas smoothing over several bins will make their errors artificially low and introduce modeling bias.

The construction of a set of radial SZE profiles from jack-knived noise maps, described in \S3.2, is used to overcome this problem. The X-ray systematic error due to uncertainties in the background modeling is treated as an amplified white noise, and random realizations of X-ray brightness profiles are obtained. Each of these X-ray profiles are then combined with one deconvolved SZE profile, and the de-projected density and temperature profiles are obtained. The scatter in each SZE profile is reduced by noise correlation, which keeps the numerical errors coming from Abel's inversion method at a minimum. Apart from density and temperature, profiles for all other cluster properties (like total mass, entropy) are obtained similarly: the scatter of the profiles measures the statistical uncertainties in each bin. The treatment of X-ray systematics as random noise is justified as the uncertainties in the current de-projected temperature values originate almost entirely from the SZE measurement. For comparison, estimating errors from a ``blind'' Monte-Carlo method treating the SZE decrement value in each bin as independent gives temperature profile errors that are on average 2-4 times higher, thus making a demonstration of the decreasing gas temperature in the cluster outskirts impossible.

The effect of SZE systematic errors on the gas temperature measurements are computed by methods described in \S3.3. The relative amplitude with respect to statistical uncertainties and the radial dependence of the SZE systematic errors are similar to those found for Abell 2163 in NBP09. That work presented tabulated uncertainty values on both $T_e$ and $n_e$. 
We ignore systematic uncertainties on gas density as it is much more robustly constrained than the gas temperature. The systematic uncertainties on $T_e$ at $r_{500}$ due to confusion with the primary CMB anisotropies is $\pm 13\%$, and at $r_{200}$ it increases to nearly twice that amount. Irrecoverable loss of the SZE signal occurs for scales larger than the path of a single bolometer, and hence can not be recovered by the transfer function (see \S5.2 in NBP09). This bias amounts to a systematic uncertainty of $+15\%$ at $r_{500}$, and roughly $+40\%$ near $r_{200}$ (these numbers are accurate only in the context of an isothermal $\beta$-model). For comparison, the statistical uncertainty in the SZE map at $r_{500}$ is about $45\%$, increasing to $\sim 100\%$ at $r_{200}$. In Fig.\ref{fig:randsys} we have shown the effect of adding systematic uncertainties (by adding in quadrature with statistical errors) on the measured gas temperature values.

\subsection{Comparison with X-ray spectral analysis}

In Fig.\ref{fig:denstemp} we also showed the de-projected X-ray spectroscopic temperature measurements for A2204 (ZF08), to provide a direct comparison between our results and those derived from X-ray analysis. There is partial overlap between these two profiles within their $1\sigma$ uncertainties, however, near  $r_{2500}$ the SZE-derived temperature is systematically higher. It is beyond the scope of this paper to present a detailed discussion on X-ray spectral analysis and its biases, but we mention the fact that for multi-temperature ICM in hot clusters the spectral analysis method can significantly underestimate (by up to 40\%) the mass-weighted gas temperature (Mazzotta et al. 2004), and this effect is expected to be stronger near cluster cool cores where the line of sight crosses many temperature components.
The low temperature value in the innermost bin from our measurement may be partially caused by the numerical uncertainty of taking derivatives at the inner edge of the profile, or APEX-SZ beam smearing. Snowden et al. (2008) considered the effect of \textit{XMM-Newton} PSF smearing in analysis of this cluster, and gave a higher value of X-ray spectroscopic temperature near $1^{\prime}$ radius.

In order to avoid added complexities from the X-ray spectral de-projection, a simpler  way  is to make a projected (i.e. two dimensional) temperature profile from our measurements using an appropriate weighting scheme. The mean weighted value of the gas temperature
along the line of sight can be computed as $T_{\mathrm{proj}} \equiv \int WT dV \ / \int W
dV$, where $T$ is the de-projected gas temperature and $W$ is the weight
function. We use two different weighting schemes: the
standard emission weight with $W = n^2 \Lambda(T)$ (using $\Lambda(T) \propto
T^{-1/6}$ as discussed earlier), and the weighting for a ``spectroscopic-like''
temperature as discussed by Mazzotta et al. (2004), using $W = n^2 T^{-3/4}$. As seen in NBP09, the results are almost identical for these two methods, and projection results only for 
the Mazzotta model are used for comparison with the X-ray data. As can be expected, the effect of projection on the radial temperature profile is small when compared to the current statistical errors.

\begin{figure}[t]
\centering
\includegraphics[width=8.5cm]{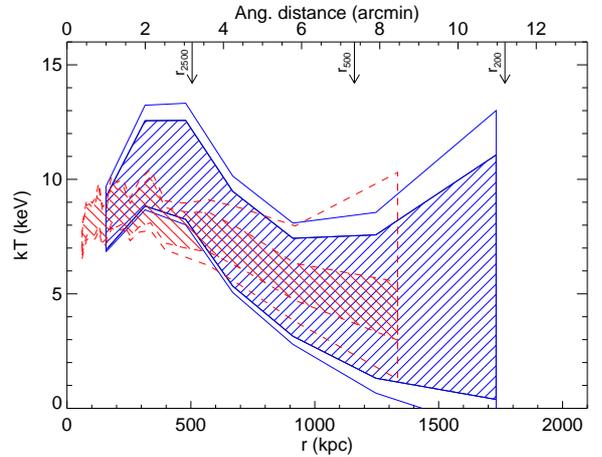}
\caption{Comparison of the projected gas temperature profile in A2204 deduced from APEX-SZ data (blue, solid boundaries) with X-ray spectral measurements from \textit{Chandra} data (red, dashed boundaries). The \textit{Chandra} spectral analysis has been re-performed taking the latest calibration update into account. The hatched regions show the $1\sigma$ statistical uncertainties in each measurement. On top of that we overplot the total uncertainties in each method combining statistical and systematic errors in quadrature. }
\label{fig:randsys}
\end{figure}

For an accurate measurement of the X-ray spectroscopic temperatures, we have 
re-analyzed two Chandra observations of A2204 (Obs IDs 6104 and 7940), 
resulting in a total exposure of 88 ks. While a temperature profile using 
these data has already been published (Sanders et al. 2009), the calibration 
update released recently (CALDB 4) was expected to have a significant effect for 
this hot cluster. Therefore, the \textit{Chandra} data was re-analyzed in the same way as described in 
Hudson et al. (2010); in addition, a correction for a 
possible difference in the cosmic X-ray background between source and 
blank sky observations was applied since we are also interested in low 
surface brightness cluster outskirts. 
The analysis with the new calibration results 
in approximately 15--20\% lower temperatures in the hot cluster regions (at $\lesssim r_{2500}$) 
as compared to Sanders et al. (2009). 

The results from the updated \textit{Chandra} spectral analysis and the projection of the SZE/X-ray 3D temperature profile are shown in Fig.\ref{fig:randsys}. The blue hatched region marks the $\pm 1\sigma$ statistical uncertainties around the mean SZE-derived temperature, and similarly the red hatched region shows the statistical uncertainties in the X-ray spectral analysis. Both results are in excellent agreement within their mutual uncertainties. But the point to note in Fig.\ref{fig:randsys} is the relative increase in the statistical and systematic errors in the \textit{Chandra} and APEX-SZ measurements of the gas temperature. The statistical and systematic errors are added in quadrature and the total uncertainties are shown on top of the statistical uncertainties (white bordered regions). At $r_{500}$ the \textit{Chandra} measurement is already dominated by systematic uncertainties due to the background modeling; beyond that radius it is impossible to put meaningful constraints on the gas temperature using the current \textit{Chandra} data. In contrast, the uncertainties on the SZE-derived temperatures are dominated by the statistical errors also at $r_{200}$. At $r_{500}$ the ratio of statistical and total systematic errors on the SZE-derived $T_{\mathrm{gas}}$ value in A2204 is roughly 2:1. 

The low systematic uncertainties in our analysis make it possible to lower the error budget on the temperature profile significantly by stacking the  SZE signal of several relaxed clusters (Basu et al., in preparation). It is true that a very long exposure will drive down the systematic uncertainties associated with the X-ray background correction, and the remaining systematic uncertainties in the flux calibration are small ($\lesssim$ 5\% for \textit{Chandra}, Vikhlinin et al. 2005). A precise comparison between the two gas temperature profiles, derived from joint SZE/X-ray analysis and X-ray spectroscopy, will be the most promising way to observationally constrain gas clumping and non-LTE effects near a cluster's virial radius. 

We mention here the recent advances made by the X-ray spectral analysis method to constrain gas temperatures out to $r_{200}$ using the \textit{Suzaku} experiment (Reiprich et al. 2009, George et al. 2009, Bautz et al. 2009). This is due to the low level of particle background in the \textit{Suzaku} orbit as compared to \textit{XMM-Newton} and \textit{Chandra}. For Abell 2204 Reiprich et al. (2009) have constrained the gas temperature near $r_{200}$ at $4.49^{+1.18}_{-0.91}$ keV, including both systematic and statistical errors. This is far superior to the current uncertainties in the APEX-SZ measurement. However, the extended PSF of \textit{Suzaku} limits its ability to spectroscopically measure the gas temperature out to the cluster virial radius to only low redshift $(z\lesssim 0.2)$ massive clusters (most of which are too extended for single-frequency APEX-SZ measurement). This also makes modeling of the gas temperature at the inner radial bins difficult. Joint SZE/X-ray temperature modeling with \textit{XMM-Newton}, \textit{Chandra} or \textit{ROSAT} data for X-ray surface brightness is therefore promising for the majority of  clusters out to high redshifts.

\subsection{Direct comparison of de-projected pressure profile with parametric models}

Applying Abel's inversion technique to the SZE map produces an unbiased and non-parametric estimate of the cluster pressure profile for a spherically symmetric system. 
This can be used to compare the usability of different parametric models, needed to extract cluster properties like $M_{200}$, from SZE or X-ray measurements made within $r_{2500}$. For example, parametric extension is unavoidable while using interferometric measurements of the SZE signal in low and intermediate redshift clusters.

\begin{figure}[t]
\centering
\includegraphics[width=8.5cm]{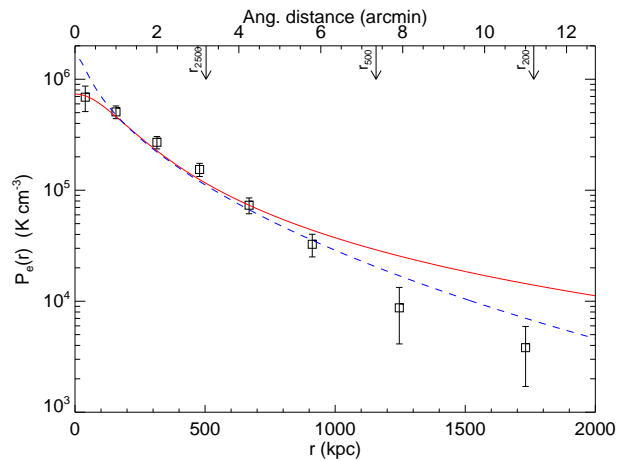}
\caption{Comparison of the de-projected pressure profile with commonly used parametric models. The data points with errors show the result of direct inversion of the SZE temperature decrement map using Abel's integral. The red (solid) line is the best fit isothermal $\beta$-model to the APEX-SZ map, fitted within $6'$ radius. The blue (dashed) line is the best fit Nagai model fitted within the same radius. The limited fitting radius was used to illustrate how well parametric modeling fitted within cluster central regions does reproduce the gas pressure in the outskirts.}
\label{fig:pressprof}
\end{figure}

The de-projected pressure profile assuming spherical symmetry for A2204 is presented in Fig.\ref{fig:pressprof}, error bars show the $1\sigma$ statistical uncertainties in the SZE measurement. We have plotted the best-fit spherical isothermal $\beta$-model and Nagai model (Nagai et al. 2007) fits on this profile. The fits are limited only to data within $6^{\prime}$ radius, to mimic an SZE observation with limited spatial extent that uses parametric model fitting to extrapolate out to the cluster virial radius. We used the Nagai profile parameters as used by the recent SZA analysis of pressure profiles (Mroczkowski et al. 2009), with parameters $(a,b,c) = (0.9, 5.0, 0.4)$. Our fitted scale radius is much larger than the predicted value of $r_p \approx r_{500}/1.3$ ($7.3^{\prime}$ in our fit), although it is strongly degenerate with the normalization factor. Similarly, the $\beta$-model parameters $[r_c,~\beta]$ are also highly degenerate. We use fit values  $r_c=1.4^{\prime}$ and $\beta=0.51$ for this comparison; setting $\beta=1$ we obtain $r_c=2.0^{\prime}$ which provides a marginally better fit to the peak SZE decrement in the deconvolved map.

The $\beta$-model is found to provide a poor extrapolated fit to the pressure profile even at $r_{500}$ (Fig.\ref{fig:pressprof}), whereas the Nagai model provides a much better fit. The two outer bins representing roughly the values at $r_{500}$ and $r_{200}$ in our analysis have $3\sigma$ upper limits at $2.25\times 10^4$ keV cm$^{-3}$ and $1.01\times 10^4$ keV cm$^{-3}$, respectively. This puts the values predicted by the Nagai model at these radii at roughly $1.7\sigma$ above our measured values, whereas the $\beta$-model predictions are at $4-5\sigma$ off. Note the limited spatial range used in this comparison study; a more accurate $\beta$-model fit for the full SZE-derived pressure profile is possible given the degeneracy of the fitting parameters.

We have also attempted to fit the de-projected pressure profile using a polytropic model for the gas, where the pressure and density are related by the relation $P(r) \propto n_e(r)^{\gamma}$. Many authors continue to use this model, e.g. Afshordi et al. (2007) use it to constrain cluster pressure profiles from WMAP data. This model is found to be too steep for the pressure profile near the cluster center, although in the outer regions ($r\gtrsim 1$ Mpc) the fit is good (using $\gamma=1.2$, e.g. Finoguenov et al. 2001). A combination of $\beta$-model in the cluster center and polytropic model in the outskirts can be used to fit the entire pressure profile, in particular to avoid the cuspiness of the Nagai model at the center. A more comprehensive analysis of the different parametric models to describe the SZE-derived pressure profile near $r_{200}$ will be discussed in a future paper.

\section{Mass and entropy profiles of the ICM}

If the ICM is in hydrostatic equilibrium (HSE) within the DM gravitational potential, the gas temperature reflects directly the depth of the potential well. The ratio between gas mass and total mass as function of radius shows the amount of baryons that is contained in the ICM. A low value of the ICM mass fraction, or a falling gas entropy profile, can indicate the existence of multi-phase ICM with non-thermal pressure support near the cluster virial radius, and physical processes hitherto unexplored in numerical cluster simulations. 

\subsection{Gas mass and total mass distribution}

The total mass, $M_{\mathrm{total}}$, is obtained by solving the hydrostatic
equilibrium equation assuming spherical symmetry (e.g. Sarazin 1988):
\begin{equation}
M_{\mathrm{total}}(<r) = - \frac{k_{\mathrm{B}}T_e(r) \ r}{G \mu m_p} \ 
\left[\frac{d \ln n_e(r)}{d \ln r} + \frac{d \ln T_e(r)}{d \ln r}\right].
\label{eq:totmass}
\end{equation} 
where $T_e(r)$ and $n_e(r)$ are the electron gas temperature and density radial profiles, and $\mu=0.62$ is the mean molecular weight per hydrogen atom, assuming primordial abundance. As can be seen, the total mass is primarily a function of the gas temperature, and only weakly dependent (through logarithmic derivatives) on the slopes of the density and temperature profiles. Therefore, the uncertainties in our temperature measurement are directly reflected in the total mass profile. The gas mass is computed directly from the de-projected density as $\rho_{\mathrm{gas}}(r) = \mu_e m_p n_e(r)$, with $\mu_e=1.17$ the mean molecular weight per electron. The gas mass fraction is simply the ratio: 
$f_{\mathrm{gas}}(<r) = M_{\mathrm{gas}}(<r)/M_{\mathrm{total}}(<r)$.

\begin{figure}[t]
\centering
\includegraphics[width=8.5cm]{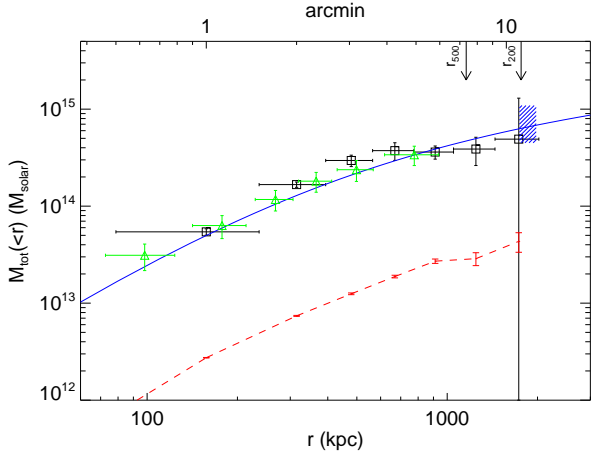}
\caption{Cumulative gas mass and total mass profiles in Abell 2204 from the SZE/X-ray joint analysis. The black data points (boxes) show the result using the de-projected temperature and density profiles, assuming hydrostatic equilibrium condition. The green data points (triangles) show the results from \textit{XMM-Newton} analysis by ZF08 under the same assumption. The blue solid line is the best fit NFW model from weak lensing analysis by Corless et al. (2009), the hatched region at $r_{200}$ indicates their quoted errors in $M_{200}$. The lower red dashed line shows the gas mass profile computed directly from the de-projected electron density.}
\label{fig:massplot}
\end{figure}

\begin{figure}[t]
\centering
\includegraphics[width=8.5cm]{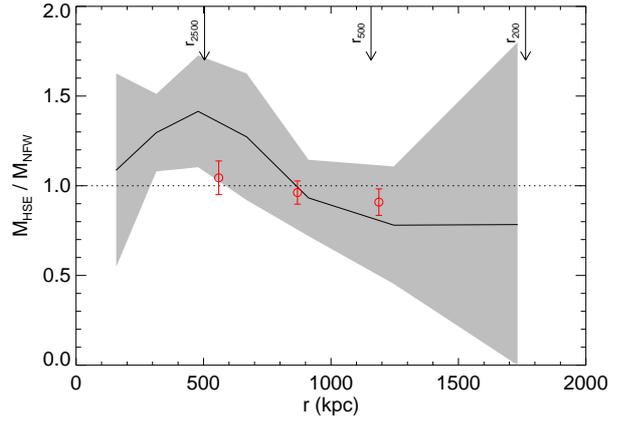}
\caption{Ratio of the total cluster mass derived from the hydrostatic equilibrium assumption, to the best fit NFW model obtained from the optical data. The cumulative total mass derived from APEX-SZ temperature measurements are consistent with the NFW model by Corless et al. (2009) under a spherical prior. For comparison, three data points from the stacking analysis for X-ray (XMM-Newton) to weak lensing (Subaru) mass ratio in 5 low redshift relaxed clusters are shown (Zhang et al. 2010).}
\label{fig:massratio}
\end{figure}

Results from the non-parametric mass modeling are shown in Fig.\ref{fig:massplot}, in comparison with results from recent X-ray and weak lensing analyses. The mass profile is in excellent agreement with the X-ray results obtained under the same assumptions of spherical symmetry and hydrostatic equilibrium (ZF08). The solid line in Fig.\ref{fig:massplot} refers to the best fit NFW model for A2204 from weak lensing analysis by Corless et al. (2009). The mean mass profile indicates a slowly rising integrated mass near $r_{500}$, resulting from the rapid fall in gas temperature near this radius, although the results are still consistent with the NFW model from Corless et al. (2009) within $1\sigma$ errors. ZF08 reports a value of $M_{500}=(5.8\pm 1.6)\times 10^{14}~M_{\odot}$ from the X-ray modeling, and the weak lensing analysis under a spherical prior gives $M_{500} = 5.3\times 10^{14}~M_{\odot}$. Our SZE/X-ray joint de-projection analysis predicts $M_{500}=(3.7\pm 1.2)\times 10^{14}~M_{\odot}$, somewhat lower than the X-ray and weak lensing results. The $1\sigma$ upper limit on $M_{200}$ from our analysis is $1.01\times 10^{15}~M_{\odot}$, again consistent with the Corless et al. (2009) maximum likelihood model prediction of  $M_{200}=0.71^{+0.38}_{-0.26}\times 10^{15}~M_{\odot}$ (shown by the blue hatched region in Fig.\ref{fig:massplot}).

To show more clearly the deviation of the non-parametric mass modeling under the HSE assumption from the weak lensing mass, we plot in Fig.\ref{fig:massratio} the ratio of the hydrostatic mass and the mass derived from the weak lensing analysis. Optical observations of A2204 are complicated by its low galactic latitude and presence of a $M_V=5.6$ star $4.3^{\prime}$ away from the center, making its shear profile in the cluster center extremely noisy (Clowe \& Schneider 2002). Thus we have used the profile from the NFW model fit to the weak lensing data instead. The ratio obtained is mostly consistent with 1 within $1\sigma$ statistical errors. The HSE assumption is expected to under-estimate the total mass by 15--20\% near the virial radius,  due to the stochastic gas motions caused by infalling matter (Nagai et al. 2007, Ameglio et al. 2009). However, the current uncertainties on the APEX-SZ measurement of a single cluster are too large to confirm any such trend. For comparison, we have shown in Fig.\ref{fig:massratio} the recent results from the stacking analysis for the $M_{\mathrm{HSE}}/M_{\mathrm{WL}}$ ratio using \textit{XMM-Newton} and \textit{Subaru} data of 5 relaxed clusters (Zhang et al. 2010). Note however, that this joint X-ray/weak-lensing stacking analysis uses the actual weak lensing shear measurements in clusters and not the best-fit NFW profiles, which should cause their ratio to be closer to unity than ours.

\begin{figure}[t]
\centering
\includegraphics[width=8.5cm]{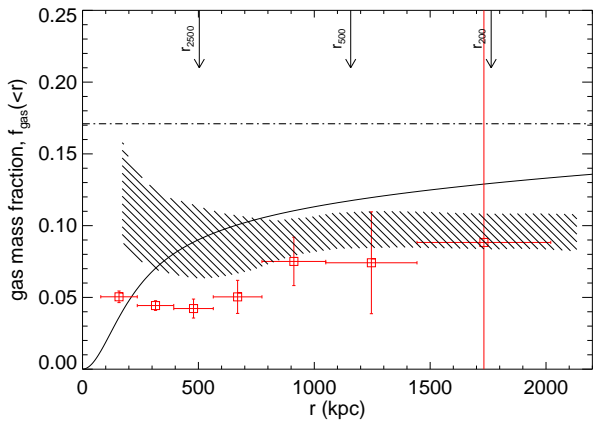}
\caption{The gas mass fraction obtained from the SZE-X/ray joint non-parametric analysis (red squares). The hatched region is the measurement from Afshordi et al. (2007) from the stacking analysis of 193 massive clusters in the WMAP 3-year data. The solid line is the prediction from the isothermal $\beta$-model fit to the SZE deconvolved map ($r_c=1.78',~ \beta=0.812$), and the horizontal dot-dashed line represents the cosmic baryon fraction from the WMAP 5-year result (Dunkley et al. 2009).}
\label{fig:gasfrac}
\end{figure}

The integrated baryon fraction of the ICM as function of radius is computed directly by dividing the gas mass by the total mass obtained from the HSE assumption. The results from our non-parametric analysis are shown in Fig.\ref{fig:gasfrac}. The cosmic baryon fraction obtained from the WMAP 5-year result ($\Omega_b/\Omega_m = 0.165\pm 0.009$, Dunkley et al. 2009) is shown in horizontal dot-dashed line. In the inner region of the cluster ($r\lesssim r_{500}$) the gas-to-mass ratio is clearly much lower than the cosmic baryon fraction, and there is an indicative trend of increasing ICM mass fraction at larger radii. Near $r_{200}$ the cumulative value of $f_{\mathrm{gas}}$ is statistically consistent with the cosmic value. Low values of gas mass fraction near cluster centers is well known from X-ray studies; Vikhlinin et al. (2006) have shown the value of $f_{\mathrm{gas}}$ at $r_{2500}$ for a sample of nearby relaxed cluster to be in the range $0.04-0.1$. It is interesting to note, however, that the gas mass fraction in Abell 2204 at $r_{2500}$ is significantly lower than previous APEX-SZ measurements of this ratio in non-relaxed clusters. The integrated $f_{\mathrm{gas}}$ values in Abell 2163 at $r\gtrsim r_{2500}$ were found to be consistent with the cosmic baryon fraction (NBP09), and HL09 measured the integrated $f_{\mathrm{gas}}$ for the Bullet cluster (1E 0657-56) within $r_{2500}$ and 1.42 Mpc to be in the range $0.18-0.22$. One obvious explanation for the higher $f_{\mathrm{gas}}$ value at the center of dynamically complex systems can be due to the fact that merging activity will most likely cause the gas to remain at the center while the dark matter halos are separated, thus causing an increase in the gas-to-mass ratio. Also the central AGN in the strong cool core cluster A2204 can be responsible for driving out the gas from the innermost region (Bhattacharya et al. 2008, Puchwein et al. 2008). 

The $f_{\mathrm{gas}}$ profile predicted from the isothermal $\beta$-model (fit to the SZE data) is shown in Fig.\ref{fig:gasfrac}. As noted in NBP09, our non-parametric modeling shows clear departure from the typical isothermal $\beta$-model prediction of $f_{\mathrm{gas}}\rightarrow 0$ at the cluster center. We have also shown for comparison the results for the stacking analysis of 193 massive clusters with $T_X > 3$ keV from the WMAP 3-year data by Afshordi et al. (2007). Note that the resolution of WMAP does not allow a direct measurement of the mean pressure profile in clusters down to $0.1r_{200}$, the hatched region is the prediction from their numerical simulations with $P_{\mathrm{gas}}>0$ prior that is consistent with the WMAP stacked measurement. The stacking signal from WMAP data predicts a higher ICM mass fraction than seen in A2204 near the center, although beyond $r_{2500}$ they are consistent with each other within $1\sigma$. The difference near the center is expected, since the sample of Afshordi et al. (2007) contains both relaxed and non-relaxed clusters, which results in a higher average $f_{\mathrm{gas}}$ value. 
If similar low gas-to-mass ratios are found consistently from SZE observations of massive relaxed clusters, then in parallel with the currently favored lower value of $\sigma_8$ parameter (Komatsu et al. 2009), this will cause significantly low cluster yields in blind SZE surveys.

\subsection{The entropy profile of the ICM}

The entropy profile can be considered a more fundamental property for analyzing the thermodynamic state of the ICM than density or temperature, as it depends directly on the history of heating and cooling within the cluster. Consequently, cluster entropy profiles have received significant attention in X-ray studies (e.g. Voit et al. 2005, Morandi \& Ettori 2007, Cavagnolo et al. 2009), but no direct measurement of entropy from SZE-derived temperatures have been done. The latter can potentially overcome the biases inherent in the X-ray spectral analysis, caused by gas clumping and substructures, and also from multiple temperature components near the cluster core. In this final part of our work we present the first SZE-derived entropy measurement in a cluster.

We adopt the standard definition of gas entropy used in the X-ray literature: $K= T_e n_e^{-2/3}$ (Ponman et al. 1999). This relates to the classical thermodynamic entropy in an ideal gas as $s = \mathrm{ln}K^{3/2} + \mathrm{constant}$. Simulations for self similar cluster models predict an entropy profile in the form of a power law: $K(r) \propto r^{1.1}$ (Tozzi \& Norman 2001, Voit et al. 2005), except at the very core of the cluster where excess entropy due to non-gravitational heating processes tend to flatten the entropy profile. This general behavior has been verified from numerous X-ray observations (e.g. Lloyd-Davies et al. 2000, Morandi \& Ettori 2007, Cavagnolo et al. 2009).

\begin{figure}[t]
\centering
\includegraphics[width=8.5cm]{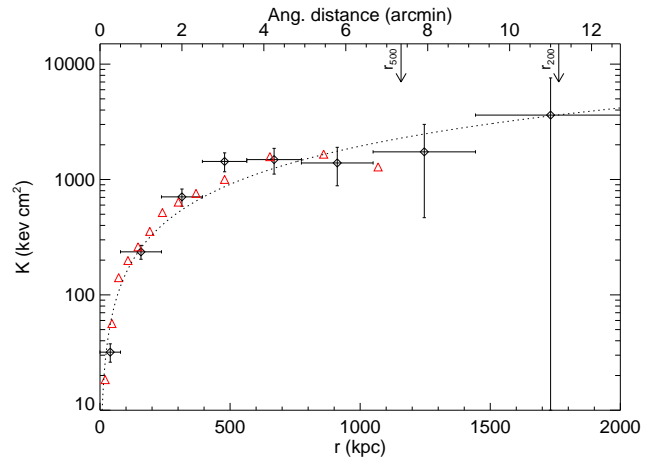}
\caption{Comparison of entropy profile in A2204 from the SZE-X-ray joint analysis, and \textit{Chandra} X-ray analysis by Sanders et al. (2008). The red triangles are from the \textit{Chandra} data, shown without errors. The dotted line is the best fit entropy profile of the form $K(r) \propto r^{1.1}$ using the data points within $8'$ radius. The present statistical errors fail to give any definitive indication for low entropy gas near $r_{200}$.}
\label{fig:entprof}
\end{figure}

The entropy profile of Abell 2204 obtained from the SZE temperature measurements is shown in Fig.\ref{fig:entprof}. Also shown are the radial entropy values obtained by Sanders et al. (2009) from the \textit{Chandra} measurements (triangles). The statistical errors in their measurements are comparable with the symbol sizes and much smaller than the present SZE measurement errors (although note that the systematic uncertainties should dominate the errors in the outer \textit{Chandra} bins, which was not shown by Sanders et al. 2009). 
The two measurements agree within the $1\sigma$ uncertainties of our analysis. The agreement between the APEX-SZ and \textit{Chandra} values within the central $1^{\prime}$ radius shows that any downward bias on the APEX-SZ value due to PSF smearing is sufficiently small. 
The dotted line in Fig.\ref{fig:entprof} is the power-law prediction from self-similar cluster models, fitted to our entropy measurement within $8'$ radius.

\begin{figure}[t]
\centering
\includegraphics[width=8.5cm]{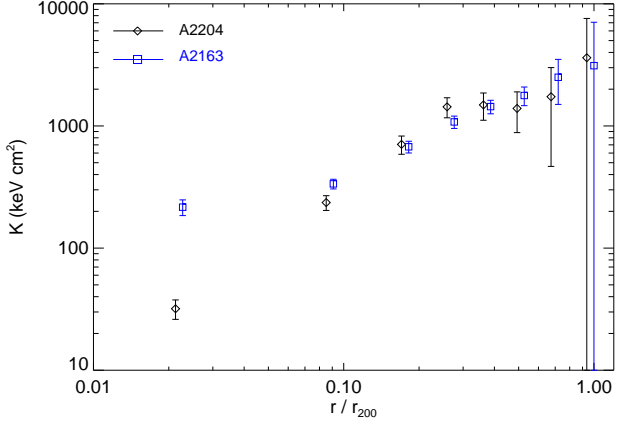}
\caption{Comparison between entropy profiles for A2204 (black, diamonds) and A2163 (blue, squares). The increased entropy value in the central region of A2163 supports the merging nature of this cluster, whereas the continually decreasing entropy towards the center in A2204 shows its dynamically relaxed state. }
\label{fig:entcomp}
\end{figure}

A flattening of the entropy profile near the virial radius of clusters has been measured only very recently from X-ray spectral analysis with \textit{Suzaku} data: for PKS 0745-191 ($z=0.10$, George et al. 2009) and Abell 1795 ($z=0.06$, Bautz et al. 2009). Note that the result of Bautz et al. (2009) was obtained by extrapolating the values within $r_{500}$. Such flattening or dropping entropy profile is another indicator of the non-thermal pressure support in the cluster outskirts. From the mean de-projected radial density and temperature profiles, we also see similar flattening for Abell 2204 near $r_{500}$ (Fig.\ref{fig:entprof}), but the statistical uncertainties in these SZE measurements are too large which makes our profile compatible with the power law scaling of the self similar cluster models.

The 1 arcmin resolution of the APEX-SZ experiment is sufficient to compare the gas entropy at the cluster cores for low and intermediate redshift massive clusters ($r_{200} \gtrsim 8'$). As mentioned before, this comparison is important in analyzing the cluster dynamical states, by comparing the extent of non-gravitational heating or cooling at the cluster core. We have used the results for Abell 2163 from the analysis in NBP09, which is a good example of a dynamically complex cluster which most likely has undergone a merging event (Maurogordato et al. 2008). The data for A2163 is analyzed with the same radial binning as for A2204 used in the current paper. The comparison of the entropy profiles between these two clusters is shown in Fig.\ref{fig:entcomp}. The difference in the central entropy values is clearly seen. Although the central bin errors are possibly under-estimated by a small amount because of neglecting PSF extension, the statistical significance of the entropy difference ($\sim 6\sigma$) is sufficiently high. Outside the central $\sim 2$ arcmin the two profiles are consistent with each other and both follow the $\propto r^{1.1}$ scaling law. It is interesting to note that the merging system A2163 shows better agreement with the power law scaling than the relaxed cluster A2204. Using a $\chi^2$ statistic to describe the goodness of fit is problematic because of the noise correlation, nevertheless, it can be used to compare the results between two clusters. 
A fit with the expected power law scaling excluding the inner- and outermost bins for A2204 gives $\chi^2/\mathrm{d.o.f.}=7.4/5$, and for A2163 it is $\chi^2/\mathrm{d.o.f.}=3.2/5$.

To conclude, we are able to demonstrate the correlation of the ``entropy floor'' with the dynamical state of a cluster (Fig.\ref{fig:entcomp}) for the first time using SZE imaging data, independent of X-ray spectral analysis. This correspondence shows the potential for the current de-projection technique, using X-ray surface brightness maps and already available SZE imaging data from multi-pixel bolometer array experiments, to select relaxed clusters from a large sample where deep X-ray observation is not available/required. This will be important, for example, for selecting relaxed clusters to constrain cosmology with the ICM baryon fraction.

\section{Conclusions}

\begin{enumerate}

\item We describe the detailed application of a direct, nonparametric de-projection method of cluster density and temperature profiles, using APEX-SZ and XMM-Newton data.  The method was presented in NBP09, the current paper builds upon the previous work by applying this technique to the well-studied relaxed cluster Abell 2204.

\item Analysis of both SZE and X-ray data have been done from their raw data sets, to create images and radial profile. In particular, we describe the creation of a set of SZE radial profiles, all consistent with the APEX-SZ measurement, to characterize the statistical uncertainties on the bin values and minimize the numerical errors in Abel's de-projection method. Our final results are dominated by the statistical uncertainties in the SZE data, the signal at $r_{200}$ is essentially an upper limit for A2204. We describe the different sources of systematic uncertainties and include them in the analysis.

\item The decreasing gas temperature in the cluster outskirts is demonstrated for the first time from SZE measurements, using a broad re-binning of the APEX-SZ (and X-ray) data. The temperature drop can be confirmed to $98\%$ confidence level. We also compare the direct de-projected pressure profile with some parametric models, and show that the Nagai profile is adequate for modeling the gas pressure, within the current statistical uncertainties in APEX-SZ imaging of a single cluster.

\item We re-perform the X-ray spectral analysis for the ICM temperature profile from publicly available \textit{Chandra} data, primarily to find the changes from the recent calibration updates (CALDB 4), but also to show the effect of systematic uncertainties due to the background modeling in the X-ray spectral analysis. A comparison with the projected temperature profile obtained from SZE data confirms that our SZE derived temperature values are much less affected by systematic uncertainties at large radii, in comparison with \textit{Chandra} and \textit{XMM-Newton}. Precise comparison between the SZE and X-ray spectroscopic measurements of the gas temperature in the cluster outskirts will be a promising method to constrain gas clumping and non-LTE effects. 

\item The integrated total mass profile is computed assuming hydrostatic equilibrium for the cluster gas. The mass profile is in excellent agreement with the recent X-ray and weak lensing analyses. Our model prediction for $M_{500}$ is $(2.6\pm 2.2) \times 10^{14}~h^{-1} M_{\mathrm{\odot}}$. This is somewhat lower than the X-ray and lensing results but consistent within $1\sigma$ errors. The upper limit on $M_{200}$ from our analysis is in good agreement with the published NFW model fit from weak lensing analysis of A2204.

\item The ICM mass fraction as function of radius is computed using the non-parametric modeling, and found to be below 0.1 within $r_{500}$. The low $f_{\mathrm{gas}}$ value in A2204 in the cluster center is in contrast with the previous APEX-SZ measurement of this ratio for Abell 2163 and the Bullet cluster.

\item We compute the ICM entropy profile from SZE/X-ray joint analysis and confirm the general agreement with the self-similar cluster model predictions within the present statistical uncertainties. The significance of the APEX-SZ measurement of A2204 is not sufficiently high at $r\gtrsim r_{500}$ to constrain the slope of the entropy profile in the cluster outskirts. 

\item We compare the entropy profiles of Abell 2204 (relaxed) and Abell 2163 (merging system), using the same non-parametric SZE/X-ray de-projection and radial binning, and find a clear entropy difference in their central 200 kpc. This corresponds to the different dynamical states of these two clusters and seen for the first time from SZE derived $T_{\mathrm{gas}}$ measurement.

\end{enumerate}

\begin{acknowledgements}

We appreciate the comments from the anonymous referee which have improved the discussion on the future applicability of this method.
We thank the APEX staff for their assistance during APEX-SZ observations. This work has been partially supported by the DFG Priority Programme 1177 and Transregio Programme TR33.   
APEX is a collaboration between the Max-Planck-Institut f\"ur Radioastronomie, the European Southern Observatory, and the Onsala Space Observatory. APEX-SZ is funded by the National Science Foundation under Grant No. AST-0138348. 
The XMM-Newton project is an ESA Science Mission with instruments and
contributions directly funded by ESA Member States and the USA
(NASA). The XMM-Newton project is supported by the Bundesministerium
f\"ur Wirtschaft und Technologie/Deutsches Zentrum f\"ur Luft- und
Raumfahrt (BMWI/DLR, FKZ 50 OX 0001) and the Max-Planck Society.  
KB acknowledges Hans B\"ohringer for discussion and reading the manuscript. 
YYZ and THR acknowledges support by the DFG through Emmy Noether Research Grant
RE\,1462/2 and by the BMBF/DLR grant No.\,50\,OR\,0601. 
MN and FPN acknowledges support for this research through 
the International Max Planck Research School (IMPRS) for Radio and Infrared
Astronomy at the Universities of Bonn and Cologne. 

\end{acknowledgements}

\end{document}